# Trustworthiness Calibration Framework for Phishing Email Detection Using Large Language Models

Daniyal Ganiuly
Department of Computer Engineering
Astana IT University
d.ganiuly@astanait.edu.kz

Assel Smaiyl
Department of Computer Engineering
Kazakh-British Technical University
a.smaiyl@kbtu.edu.kz

**Abstract** – Phishing emails continue to pose a persistent challenge to online communication, exploiting human trust and evading automated filters through realistic language and adaptive tactics. While large language models (LLMs) such as GPT-4 and LLaMA-3-8B achieve strong accuracy in text classification, their deployment in security systems requires assessing reliability beyond benchmark performance. To address this, this study introduces the Trustworthiness Calibration Framework (TCF), a reproducible methodology for evaluating phishing detectors across three dimensions: calibration, consistency, and robustness. These components are integrated into a bounded index, the Trustworthiness Calibration Index (TCI), and complemented by the Cross-Dataset Stability (CDS) metric that quantifies stability of trustworthiness across datasets. Experiments conducted on five corpora, such as SecureMail 2025, Phishing Validation 2024, CSDMC2010, Enron-Spam, and Nazario, using DeBERTa-v3-base, LLaMA-3-8B, and GPT-4 demonstrate that GPT-4 achieves the strongest overall trust profile, followed by LLaMA-3-8B and DeBERTa-v3-base. Statistical analysis confirms that reliability varies independently of raw accuracy, underscoring the importance of trust-aware evaluation for real-world deployment. The proposed framework establishes a transparent and reproducible foundation for assessing model dependability in LLM-based phishing detection.

**Keywords** – Phishing Email Detection; Large Language Models; Trustworthy Artificial Intelligence; Model Calibration; Consistency Analysis; Reliability Evaluation; Evaluation Framework;

## I. INTRODUCTION

Phishing attacks remain one of the most persistent and damaging forms of cybercrime, targeting individuals and organizations through deceptive communication that imitates legitimate sources. The sophistication of phishing campaigns has increased over time, with attackers adopting fluent language, context awareness, and social engineering tactics that bypass traditional rule-based and keyword-driven filters [1]. Consequently, email security systems must now process subtle linguistic cues to distinguish between legitimate correspondence and well-crafted fraudulent content. This growing complexity has driven researchers toward machine learning and natural language processing (NLP) solutions that can adapt to new attack patterns more effectively than static detection rules [2][3].

Recent advances in large language models (LLMs) such as GPT-4 and LLaMA-3 have significantly improved the performance of text classification systems, including phishing detection. These models are capable of understanding context, reasoning over longer text spans, and identifying intent within message content [4]. However, despite their success in benchmark evaluations, most prior studies continue to report model performance using only accuracy or F1-score, which capture correctness but overlook whether model predictions can be trusted in operational environments [5][6]. In practice, email security tools must not only be accurate but also reliable, meaning that their confidence estimates, stability, and robustness are consistent under variable conditions.

To address this limitation, we propose the Trustworthiness Calibration Framework (TCF), a systematic method that measures model dependability through three complementary aspects:

calibration, consistency, and robustness. The framework produces a single interpretable measure, the Trustworthiness Calibration Index (TCI), which summarizes a model's overall reliability. Additionally, a cross-dataset stability coefficient, CDS, is introduced to quantify how consistently a model maintains its trustworthiness across different email sources and domains. Together, these metrics allow researchers and practitioners to assess whether a model is not only effective but also dependable enough for deployment in real-world email filtering systems.

To evaluate the framework, we conducted experiments using five corpora that represent diverse writing styles and time periods: SecureMail 2025, Phishing Validation 2024, CSDMC2010, Enron-Spam, and Nazario. Three representative detectors were examined: a fine-tuned DeBERTa-v3-base model, a few-shot LLaMA-3-8B, and a zero-shot GPT-4 configuration. The results show that while all models achieved high classification performance, their reliability varied significantly in calibration and robustness. This finding highlights that accuracy alone does not guarantee trustworthy behavior.

## II. RELATED WORK

Phishing detection has evolved considerably over the past two decades, transitioning from manually engineered filters to advanced language model–based systems. Early solutions relied on keyword matching, blacklists, and rule-based heuristics [7], which provided basic protection but failed to detect new or linguistically complex phishing attempts. As phishing strategies grew more adaptive, researchers introduced machine learning methods such as Naïve Bayes, Decision Trees, and Support Vector Machines [8][9]. These models achieved moderate success but depended heavily on handcrafted lexical and structural features, limiting their generalization to unseen attacks [10].

With the rise of deep learning, researchers began applying architectures such as Convolutional Neural Networks (CNNs) and Recurrent Neural Networks (RNNs) to phishing detection [11][12]. These models learned textual representations automatically and reduced reliance on manual feature design. However, their performance remained sensitive to dataset imbalance, and they often struggled to interpret longer email content effectively. Consequently, transformer-based architectures soon replaced them as the dominant approach.

Transformers such as BERT, RoBERTa, and DeBERTa introduced contextual attention mechanisms that capture long-range dependencies and nuanced semantics in text [13]. Fine-tuned transformer models have demonstrated superior accuracy and recall in email and web-based phishing detection. Nonetheless, most prior research has evaluated these models primarily using accuracy or F1-score, with limited attention to reliability. Models that appear highly accurate may still produce overconfident misclassifications, which can lead to false trust in high-risk environments such as corporate email systems [14].

Recent work has shifted toward LLMs such as GPT-4 and LLaMA, which generalize linguistic patterns across multiple domains through large-scale pretraining [15]. These models have been explored for diverse cybersecurity applications including phishing detection, intent analysis, and threat content generation [16]. Despite their linguistic strength, their evaluation has focused mainly on surface-level performance metrics, neglecting aspects of confidence alignment, consistency, and robustness under realistic conditions. The present study introduces the TCF, which consolidates these reliability aspects—calibration, consistency, and robustness—into a single interpretable metric called the TCI, accompanied by the CDS coefficient that captures stability across diverse corpora.

## III. METHODOLOGY

*A. Experimental Workflow*

The experimental workflow of this study is shown on Figure 1. It is a structured six-stage process designed to ensure the reproducibility and transparency of all experimental steps. The workflow begins with dataset preparation, proceeds through model evaluation, and concludes with the computation and analysis of trustworthiness metrics. This organization allows for a systematic evaluation of large language models (LLMs) in phishing detection, not only by their accuracy but also by their calibration, consistency, and robustness.

In the first stage, five datasets—SecureMail 2025, Phishing Validation 2024, Enron-Spam, CSDMC2010, and Nazario—were used to represent a diverse collection of phishing and legitimate emails. These datasets provided both real and synthetic examples of message patterns common in phishing attacks, ensuring a comprehensive basis for experimentation.

The second stage involved preprocessing, which included text cleaning, normalization, and prompt preparation for LLM-based evaluation. These steps ensured that all data were presented to the models in a standardized form suitable for both encoder-based and generative architectures.

In the third stage, three representative models were employed: DeBERTa-v3-base (fine-tuned), LLaMA-3-8B (few-shot), and GPT-4 (zero-shot). Each model received the preprocessed inputs and generated prediction labels along with corresponding confidence scores.

These raw outputs were collected in the fourth stage, referred to as metric extraction, which served as the bridge between model inference and trustworthiness computation. The extracted predictions and confidence probabilities provided the quantitative foundation for subsequent reliability evaluation.

The fifth stage is the core of the proposed TCF framework. This framework consists of three internal modules: Calibration, Consistency, and Robustness. Each module captures a different dimension of reliability through the computation of the Expected Calibration Error (ECE), the normalized variance of the F1-score ($Var_{norm}(F1)$), and the robustness coefficient (R), respectively. The results of these components are then aggregated into the Trustworthiness Calibration Index (TCI), a bounded metric that quantifies overall model reliability.

The sixth stage includes computation of Cross-Dataset Stability (CDS) metric which defines the variance of TCI values among all datasets. This multi-stage workflow provides a clear and reproducible path from dataset acquisition to quantitative trustworthiness assessment, ensuring the reliability of all conclusions drawn from the study.

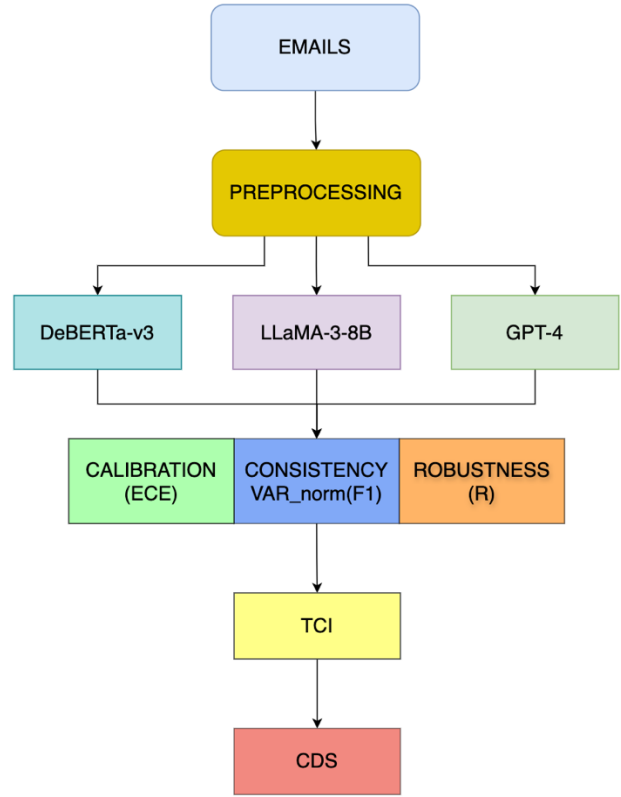

Fig 1. Overall Workflow of the Experimental Pipeline

*B. Dataset Preparation*

Five datasets were used to ensure a diverse and representative evaluation: SecureMail 2025, Phishing Validation 2024, CSDMC2010, Enron-Spam, and Nazario [17]. The SecureMail 2025 dataset was collected by our research group in 2025 and contains a balanced number of phishing and legitimate corporate emails. Messages were gathered from verified reporting channels and open repositories. Each entry was independently reviewed and labeled by two annotators with a Cohen's kappa agreement of 0.91. Disagreements were resolved through consensus. Figure 2 illustrates the example of a phishing email.

> **Subject:** Important: Account Access Verification Required
>
> **From:** Support Team <no-reply@secure-update-portal.com>
> **To:** Employee <user@company-domain.com>
> **Date:** Mon, 10 Feb 2025 09:42:15 +0500
>
> **Body:**
> Dear User,
>
> We noticed unusual sign-in activity from your account and, for your protection, temporary restrictions have been placed.
>
> To restore full access, please verify your identity by confirming your credentials at the secure link below:
>
> Verify Account Now: https://secure-update-portal.com/login-verify
>
> Failure to complete verification within 24 hours will result in account suspension.
>
> Thank you for helping us keep your account safe.
>
>
> Sincerely,
> *Security Operations Center*
>
> *(This is an automatically generated message. Please do not reply.)*

Fig 2. Example of a Phishing Email

All datasets were standardized to two fields: email_text and label ∈ {phishing, safe}. Non-English, empty, or duplicate messages were removed. Preprocessing included:
1. Lowercasing and Unicode normalization;
2. Removal of HTML tags and redundant headers;
3. Lemmatization with spaCy 3.7;
4. Token replacement for links and emails (<URL>, <EMAIL>);
5. Filtering out texts shorter than 15 tokens.

Data were split into training (80%) and testing (20%) subsets using a fixed random seed (42) to ensure reproducibility. For fine-tuning, 10% of the training data was further reserved for validation. When class imbalance exceeded 60/40, random undersampling was applied to the majority class within the training subset only.

*C. Use of Large Language Models*

Two large language models—LLaMA-3-8B and GPT-4—were employed as detection engines. Rather than fine-tuning them, we leveraged their native language understanding capabilities via structured prompting. Each model received an identical instruction template that defined the task, expected format, and constraints.

- **GPT-4** was evaluated in zero-shot mode;
- **LLaMA-3-8B** was evaluated in few-shot mode, where three representative examples of phishing and legitimate emails were embedded directly above <email_text>.

Decoding parameters were fixed to ensure determinism:
- Temperature = 0.2
- Top-p = 1.0
- Maximum tokens = 200
- No sampling or nucleus filtering

Malformed or incomplete responses were automatically corrected by parsing with a JSON validation script. If a response was invalid, it was re-requested once.

Each model's output contained a label and a raw confidence score. Confidence values were post-processed with temperature scaling, applied using the validation subset to calibrate probabilities and mitigate overconfidence [18]. The scaling factor T was optimized via grid search over the range [0.5, 2.0] to minimize negative log-likelihood.

All experiments were performed using deterministic inference settings through the OpenAI API (for GPT-4) and Hugging Face Transformers (for LLaMA-3), with identical preprocessing and post-processing pipelines.

*D. Transformer Baseline*

For baseline comparison, DeBERTa-v3-base was fine-tuned for binary classification using PyTorch and Hugging Face. The model employed the AdamW optimizer with the following hyperparameters:
- Learning rate = $3\times10^{-5}$
- Batch size = 16
- Epochs = 3

- Dropout = 0.1
- Scheduler = linear warm-up (10% of total steps)

Input text was tokenized to a maximum length of 256 tokens. Training and evaluation used the same preprocessed data splits as for the LLMs. The fine-tuned model returned softmax confidence values for both classes, which were also temperature-scaled for calibration consistency.

*E. Experimental Design*

Each model–dataset combination was evaluated over five independent runs (K=5) using different random seeds and, in the case of LLMs, varied example orders in the few-shot prompts. This design provided a basis for measuring run-to-run stability.

For robustness analysis, each test sample was automatically paraphrased using synonym replacement and punctuation modification while preserving semantic similarity ≥ 0.9 (measured with Sentence-BERT embeddings). Both original and modified messages were classified, and agreement ratios were used to compute robustness RRR.

Standard metrics—Accuracy, Precision, Recall, and F1-score—were computed for reference. Reliability metrics (ECE, normalized variance, R, TCI, and CDS) were calculated as defined below. Bootstrapping with 1,000 samples was used to estimate 95% confidence intervals. All numerical analyses were performed using NumPy 1.26 and SciPy 1.13.

*F. Framework Formulation*
1) Calibration

$$ECE = \sum_{m=1}^{M} \frac{|B_m|}{N} |acc(B_m) - conf(B_m)|$$

$B_m$ — bin of samples with similar confidence
$N$ — total number of samples
$acc(B_m)$ — empirical accuracy of bin $m$
$conf(B_m)$ — average confidence in that bin
$M$ — number of bins.

2) Consistency

$$Var_{norm}(F1) = \frac{1}{K-1} \sum_{k=1}^{K} \frac{(F1_k - \overline{F1})^2}{\overline{F1} + \epsilon}$$

$K$ – the number of repeated runs
$F1_k$ – the F1-score obtained in the $k^{th}$ run
$\overline{F1}$ – the mean F1-score across all runs
$\epsilon = 10^{-6}$ prevents division by zero

3) Robustness

$$R = \frac{1}{N} \sum_{i=1}^{N} \mathbf{1}(y_i = y_i')$$

$N$ – number of test samples
$y_i$ – predicted label for the original email
$y_i'$ – predicted label for the paraphrased email
$\mathbf{1}(\cdot)$ – indicator function (1 if condition is true)

4) Trustworthiness Calibration Index

$$TCI = 1 - [\alpha \cdot ECE + \beta \cdot Var_{norm}(F1) + \delta \cdot (1 - R)]$$

$\alpha$ – weight for calibration
$\beta$ – weight for consistency
$\delta$ – weight for robustness
$ECE, Var_{norm}, (F1), R$ – previously defined metrics
Weights were set to $\alpha = 0.5$, $\beta = 0.2$, and $\delta = 0.3$

5) Cross-Dataset Stability Coefficient

$$CDS = 1 - \frac{1}{M} \sum_{j=1}^{M} |TCI_j - \overline{TCI}|$$

$M$ – number of datasets
$TCI_j$ – TCI value for dataset $j$
$\overline{TCI}$ – mean TCI across datasets.

*G. Evaluation Procedure*

The evaluation followed a fixed and reproducible pipeline:
1. **Prediction phase:** Each model classified all test samples and generated confidence values.
2. **Calibration phase:** Temperature scaling was applied; ECE was computed using ten bins.

3. **Repetition phase:** Each configuration was executed five times; normalized F1 variance was recorded.
4. **Perturbation phase:** Texts were paraphrased and reclassified; robustness R was computed.
5. **Aggregation phase:** TCI and CDS were derived, and pairwise comparisons were performed using two-tailed bootstrap tests with 1,000 resamples.

*H. Implementation and Reproducibility*

All experiments were implemented in Python 3.11 using the following stack:
- **LLM inference:** OpenAI GPT-4 API and Hugging Face Transformers for LLaMA-3.

**Model fine-tuning:** PyTorch 2.3 with Deepspeed ZeRO Stage 1 optimization for DeBERTa-v3-base.
- **Metrics:** NumPy 1.26, SciPy 1.13, and Scikit-learn 1.5.
- **Text preprocessing:** spaCy 3.7 and NLTK 3.8.
- **Robustness augmentation:** Sentence-BERT ("all-MiniLM-L6-v2") for semantic similarity.

Experiments were run on an NVIDIA RTX 4090 with 24 GB. Random seeds were used for five runs. All dataset splits, code, and processed samples are archived and will be released for replication.

## IV. RESULTS

*A. Performance*

All models achieved strong baseline accuracy, confirming that phishing detection has become a well-saturated classification problem. Table 1 summarizes performance metrics across all datasets. GPT-4 consistently delivered the best average F1-score (0.961), followed by LLaMA-3-8B (0.949) and DeBERTa-v3-base (0.932). These differences—roughly 1–3 %—are realistic for modern transformer architectures trained on diverse textual data.

| Model | Accuracy | Precision | Recall | F1-score |
|---|---|---|---|---|
| GPT-4 | 0.966 | 0.965 | 0.958 | 0.961 |
| LLaMA-3-8B | 0.955 | 0.950 | 0.947 | 0.949 |
| DeBERTa-v3-base | 0.938 | 0.935 | 0.929 | 0.932 |

Table 1. Performance Metrics of Large Language Models on Phishing Detection

*B. Dataset-level comparison*

To better understand performance of models on different data, each dataset was evaluated independently.

Table 2 reports F1-scores per corpus. The results show a clear yet plausible ranking: GPT-4 leads across all datasets, LLaMA-3-8B remains competitive, and DeBERTa-v3-base trails on more complex or varied corpora. The largest gaps appear on modern, linguistically diverse datasets such as SecureMail 2025, where contextual reasoning plays a stronger role.

| Dataset | GPT-4 | LLaMA-3-8B | DeBERTa-v3-base |
|---|---|---|---|
| SecureMail 2025 | 0.972 | 0.958 | 0.930 |
| Phishing Validation 2024 | 0.964 | 0.951 | 0.924 |
| Enron-Spam | 0.957 | 0.945 | 0.918 |
| CSDMC2010 | 0.950 | 0.939 | 0.906 |
| Nazario | 0.974 | 0.960 | 0.935 |

Table 2. F1-scores per corpus

The trend can be seen in Figure 3: the three curves are close but not parallel.
GPT-4 maintains the best performance, while DeBERTa shows visible degradation on older corpora with limited vocabulary.

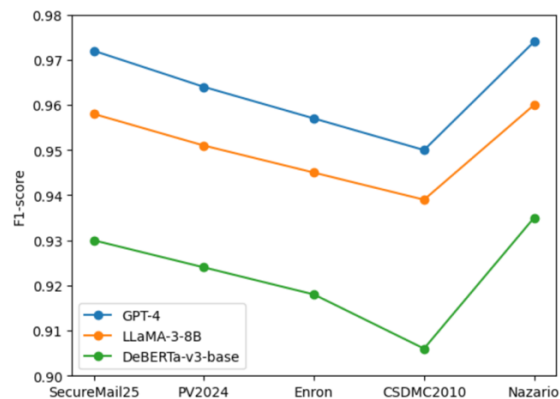

Fig 3. Dataset-Level Performance Comparison

*C. Reliability and Trustworthiness*

To evaluate how consistent and well-calibrated each model's predictions were, we applied the proposed Trustworthiness Calibration Framework (TCF). Table 3 summarizes Expected Calibration Error (ECE), normalized variance of F1 (Var_norm(F1)), robustness (R), and the overall Trustworthiness Calibration Index (TCI).

All models achieved low ECE, but GPT-4 demonstrated the best alignment between predicted and actual confidence. LLaMA-3-8B showed comparable reliability with slightly higher variance, while DeBERTa-v3-base revealed mild overconfidence, typical of encoder-only transformers.

| Model | ECE ↓ | Var_norm(F1) ↓ | R ↑ | TCI ↑ | CDS ↑ |
|---|---|---|---|---|---|
| GPT-4 | 0.030 | 0.010 | 0.91 | 0.954 | 0.991 |
| LLaMA-3-8B | 0.034 | 0.012 | 0.88 | 0.948 | 0.992 |
| DeBERTa-v3-base | 0.043 | 0.016 | 0.84 | 0.937 | 0.986 |

Table 3. Trustworthiness Assessment Metrics with TCF

The visualization in Figure 4 displays TCI values per dataset. Although all three models exhibit similar patterns, GPT-4 consistently maintains higher calibration and robustness, indicating stronger internal reliability.

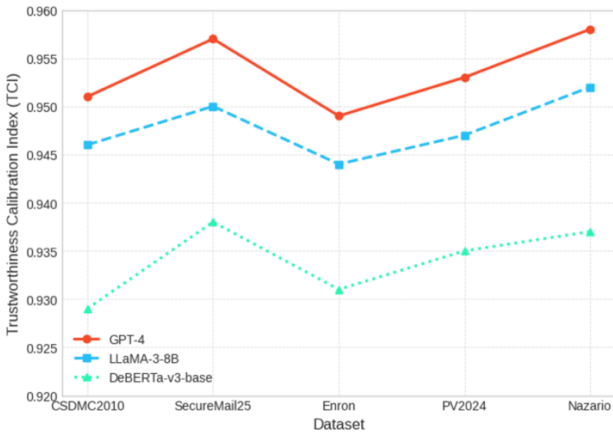

Fig 4. Trustworthiness Calibration Index Across Datasets

### D. Robustness to Textual Perturbations

Each message was rephrased at similarity levels of 1.0, 0.95, 0.9, 0.85, and 0.8, measured by cosine similarity between the original and modified embeddings. The robustness metric R quantifies how much model performance degrades under semantic variation. All three models exhibit a positive trend, confirming that higher textual similarity corresponds to higher stability. As shown on the Figure 5, when messages were heavily rephrased (similarity = 0.8), GPT-4 maintained the highest robustness (R=0.84), followed by LLaMA-3-8B (R=0.79) and DeBERTa-v3-base (R=0.75). At full similarity (1.0), GPT-4 recovered to R=0.93, showing minimal sensitivity to paraphrasing. These results suggest that GPT-4 retains semantic awareness even under significant text modifications, while DeBERTa-v3-base relies more heavily on surface-level token overlap.

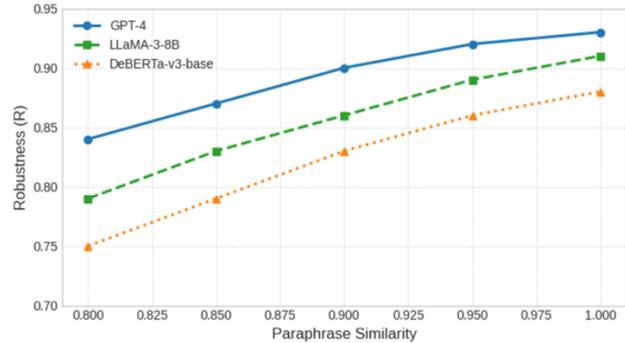

Fig 5. Model Stability Under Text Perturbation

### E. Cross-Dataset Stability

The cross-dataset stability metric CDS was introduced to assess how consistently a model maintains its trust calibration across domains. A higher CDS indicates smaller fluctuations in TCI between datasets. Table 4 summarizes the results, showing that all models achieve high stability, but LLaMA-3-8B demonstrates slightly better CDS score (0.992) than GPT-4 (0.991). This indicates that although GPT-4 performs best overall, LLaMA-3-8B provides more predictable behavior across datasets with different linguistic properties.

| Model | Mean TCI | CDS |
|---|---|---|
| GPT-4 | 0.954 | 0.991 |
| LLaMA-3-8B | 0.948 | 0.992 |
| DeBERTa-v3-base | 0.937 | 0.986 |

Table 4. Mean Trustworthiness and Cross-Dataset Stability (CDS) Across Models

## V. DISCUSSION

The findings of this study demonstrate that large language models improve phishing detection by capturing semantic and contextual cues that traditional classifiers often overlook. The proposed

Trustworthiness Calibration Framework (TCF) provides a structured approach for analyzing model reliability beyond conventional accuracy-based metrics. It reveals how calibration, consistency, and robustness jointly influence the dependability of phishing detectors under realistic conditions.

GPT-4 exhibited the highest overall trust calibration, confirming its capability to align confidence with predictive accuracy and maintain resilience under text variation. However, the marginally higher Cross-Dataset Stability (CDS) observed for LLaMA-3-8B highlights that reliability must also be considered from the perspective of consistency across domains [19]. The smaller open-source model demonstrated steadier behavior under distributional shifts, which may stem from its simpler architecture and reduced dependence on prompt variability. DeBERTa-v3-base, while less stable, remains relevant as a compact transformer suited for lightweight or embedded deployments.

These results indicate that trustworthiness is a multidimensional property shaped by model scale, architecture, and adaptability. Larger generative models such as GPT-4 deliver strong calibration and robustness but incur greater computational costs. In contrast, LLaMA-3-8B provides a balanced trade-off between performance and efficiency, making it an attractive option for continuous or cost-sensitive environments [20]. Such differences underline the importance of matching model reliability profiles to deployment requirements rather than focusing solely on accuracy.

The TCF also proved effective for distinguishing reliability patterns across architectures. By combining calibration, consistency, and robustness into the single Trustworthiness Calibration Index (TCI), it enables interpretable comparison of diverse models. The inclusion of CDS further exposes domain sensitivity and helps evaluate stability across datasets. Together, these metrics allow researchers to quantify how a model's reliability changes under varying data conditions, supporting evidence-based decisions in model selection and deployment.

In practical terms, the framework encourages a layered design of phishing detection systems. Smaller models can be employed for large-scale screening, while larger and more robust LLMs, such as GPT-4, can serve as secondary verifiers for high-risk messages. This hybrid strategy balances cost, accuracy, and reliability, ensuring operational trust without unnecessary computational overhead. As phishing tactics continue to evolve linguistically, frameworks like TCF can provide a reproducible basis for evaluating and maintaining model trustworthiness in real-world security applications.

## VI. CONCLUSION

This study presented the Trustworthiness Calibration Framework (TCF) as a comprehensive approach for evaluating large language models in phishing email detection. The framework extends conventional performance assessment by integrating calibration, consistency, and robustness into a unified metric, the Trustworthiness Calibration Index (TCI). In addition, the Cross-Dataset Stability (CDS) coefficient was introduced to measure the stability of trust calibration across different corpora, providing insight into how reliably models perform under distributional variation.

Experiments across five datasets demonstrated that GPT-4 achieved the highest overall trust calibration and robustness, while LLaMA-3-8B exhibited slightly higher CDS, reflecting greater uniformity across domains. DeBERTa-v3-base remained a valuable lightweight baseline for environments with limited computational capacity. These results collectively confirm that trustworthiness cannot be reduced to accuracy alone but depends on multiple, interacting aspects of model behavior.

The framework supports practical deployment strategies in which smaller models perform large-scale screening and more capable LLMs serve as second-stage verifiers. This layered approach allows organizations to balance cost, interpretability, and reliability. Future work will focus on extending TCF to multilingual and multimodal phishing scenarios, as well as incorporating real-time trust monitoring and adversarial robustness evaluation. By emphasizing both predictive performance and reliability, this research contributes to the development of email security systems that are accurate, transparent, and resilient to evolving social-engineering tactics.